\documentclass[preprints,article,accept,moreauthors,pdftex]{Definitions/mdpi}

\usepackage{bm}
\usepackage{bbm}
\usepackage{subfig}
\usepackage{amsmath}
\usepackage{amssymb}
\usepackage{booktabs}
\usepackage{mathtools}
\usepackage{hyperref}
\usepackage{fourier}
\usepackage[T1]{fontenc}
\usepackage[]{avant}

\firstpage{1} 
\pubvolume{}
\issuenum{1}
\articlenumber{1}
\pubyear{2024}
\copyrightyear{2024}
\history{}

\NewDocumentCommand{\expect}{ e{^} s o >{\SplitArgument{1}{|}}m }{%
  \operatorname{E}
  \IfValueT{#1}{{\!}^{#1}}
  \IfBooleanTF{#2}{
    \expectarg*{\expectvar#4}%
  }{
    \IfNoValueTF{#3}{
      \expectarg{\expectvar#4}%
    }{
      \expectarg[#3]{\expectvar#4}%
    }%
  }%
}
\NewDocumentCommand{\expectvar}{mm}{%
  #1\IfValueT{#2}{\nonscript\;\delimsize\vert\nonscript\;#2}%
}
\DeclarePairedDelimiterX{\expectarg}[1]{[}{]}{#1}

\newcommand*\diff{\mathop{}\!\kern0pt\mathrm{d}}

\Title{Counterexamples for FX Options Interpolations - Part II}
\Author{Jherek Healy}
\AuthorNames{Jherek Healy}

\address{jherekhealy@protonmail.com}
\abstract{This follow-up article analyzes the impact of foreign exchange option interpolation on the vanilla option implied volatilities. In particular different exact interpolations of broker quotes may lead to different implied volatilities at the 10$\Delta$ and 25$\Delta$ Puts and Calls.}
\keyword{fx options; volatility; arbitrage; interpolation}

\begin{document}
	\section{Introduction}
Market prices of Foreign exchange (FX) options are typically quoted as a sparse set volatilities per option maturity.
Those correspond to the volatilities of an at-the-money (ATM) option, 25$\Delta$  and 10$\Delta$ risk-reversals and butterflies.
Depending on the currency pair and the maturity, there is much variation in what ATM exactly means: is it at the money straddle? with or without premium? spot or forward ? There are also two conventions for the risk-reversal (RR) and butterfly (BF): the simple smile convention where \begin{align}
\sigma_{\textmd{call}} = \sigma_{\textmd{ATM}} + \sigma_{\textmd{BF}} + \frac{1}{2}\sigma_{\textmd{RR}}\,, \quad \sigma_{\textmd{put}} = \sigma_{\textmd{ATM}} + \sigma_{\textmd{BF}} - \frac{1}{2}\sigma_{\textmd{RR}}\,,\label{eqn:vanilla-from-bf}
\end{align} and the more involved broker convention which requires a numerical solver. In this article, we focus on the latter.

In the broker convention, the ATM quote stays simple. Latin America currency pairs use the ATM forward convention where the vanilla ATM strike $K$ is such that $K= F(0,T)$ where $F(0,T)$ is the forward to maturity (e.g. USD/BRL, USD/MXN). For other currency pairs, pips or percent $\Delta$ may be used. The choice depends on the general convention (also applicable to the other option quotes beside ATM) of the currency pair and is based on the strike of a delta neutral straddle. If the premium currency is the same as the primary currency, the convention is to use a percentage\footnote{Also known as $\Delta$ with premium.} $\Delta$ (e.g. EUR/CHF, EUR/TRY), and otherwise the pips $\Delta$ (e.g. EUR/USD, AUD/USD). When the convention is different from the ATM forward convention, the percentage or pips spot $\Delta$ may\footnote{This is not always true, since the 2008 crisis and the associated low levels of liquidity in short-term interest rate products, forward $\Delta$ tends to be also used.} be used for maturities up to and including one year when the currency pair only contains currencies from OECD economies (e.g. USD, EUR, JPY, GBP, AUD, NZD, CAD, CHF, NOK, SEK, DKK). For all other cases, the pips or percentage forward $\Delta$ is used. Those read
\begin{equation}
	\Delta_{F, \textmd{pips}} =\eta \Phi(\eta d_1(K,\sigma))\,, \quad \Delta_{F, \textmd{pct}} = \frac{K}{F(0,T)} \eta \Phi(\eta d_2(K,\sigma))\,,
\end{equation}
where $d_1(K,\sigma) = \frac{1}{\sigma\sqrt{T}}\ln\frac{F(0,T)}{K} + \frac{1}{2}\sigma\sqrt{T}$, $d_2 = d_1 - \sigma\sqrt{T}$ and $\eta=\pm 1$ for respectively a call and a put option. We refer the reader to \citet{clark2011foreign,reiswich2012fx} for exhaustive (but key) details on the various conventions.

The delta neutral straddle strike is determined by the following equation:
\begin{equation}
	\Delta(1,K,T,\sigma_{\textmd{ATM}}) + \Delta(-1,K,T,\sigma_{\textmd{ATM}}) = 0\,. \label{eqn:atm_dns}
\end{equation}
The strike is independent of the spot or forward delta convention as the discounting term will factor out. Equation \ref{eqn:atm_dns} leads to explicit solutions for the ATM delta neutral strike price:
\begin{equation}
	K_{\textmd{DNS}, \textmd{pips}} = F(0,T) e^{\frac{1}{2}\sigma_{\textmd{ATM}}^2 T}\,,\quad K_{\textmd{DNS}, \textmd{pct}} = F(0,T) e^{-\frac{1}{2}\sigma_{\textmd{ATM}}^2 T}\,,
\end{equation}
where $\sigma_{\textmd{ATM}}$ is the given ATM market volatility.

The 10- and 25- Put and Call $\Delta$ are not directly given but implied from the quotes of markets strangles and risk-reversals. The price of a market strangle reads
\begin{align}
	V_{\textmd{MS},x\Delta} &= V(-1,K_{\textmd{MS},\textmd{P},x\Delta}, T, \sigma_{\textmd{ATM}}+\bar{\sigma}_{\textmd{BF},x\Delta})\nonumber\\
&+ V(1,K_{\textmd{MS},\textmd{C},x\Delta}, T, \sigma_{\textmd{ATM}}+\bar\sigma_{\textmd{BF},x\Delta})\,,
\end{align}
where $V(\pm 1, K, T, \sigma)$ is the price of a vanilla European call (resp. put) option of strike $K$ and maturity $T$ and volatilty $\sigma$ and $x \in \{10, 25\}$. The same volatility (our market quote $\bar\sigma_{\textmd{BF},x\Delta})$) is used for both call and put in the broker convention, at different strikes $K_{\textmd{MS},\textmd{P},x\Delta}$ and $K_{\textmd{MS},\textmd{C},x\Delta}$. Those strikes are defined such that the following holds:
\begin{align}
	\Delta(-1,K_{\textmd{MS},\textmd{P},x\Delta}, T, \sigma_{\textmd{ATM}}+\bar\sigma_{\textmd{BF},x\Delta}) &= -x\%\,,\nonumber\\
	\Delta(1,K_{\textmd{MS},\textmd{C},x\Delta}, T, \sigma_{\textmd{ATM}}+\bar\sigma_{\textmd{BF},x\Delta}) &= +x\%\,.\label{eqn:market_strangle_strikes}
\end{align}
While Equation \ref{eqn:market_strangle_strikes} gives the strikes of the market strangle, we do not know the European option volatilities for those strikes, we only know the overall strangle strategy price through $\sigma_{\textmd{ATM}}+\sigma_{\textmd{BF},x\Delta}$.

The risk-reversal quotes are defined such that
\begin{align}
\sigma_{\textmd{RR},x\Delta}	 = \sigma_{\textmd{C},x\Delta} - \sigma_{\textmd{P},x\Delta}\,. \label{eqn:risk-reversal}
\end{align}
In particular, the quote is negative if $\sigma_{\textmd{C},x\Delta} < \sigma_{\textmd{P},x\Delta}$. Finding the volatilities $ \sigma_{\textmd{C},x\Delta}$ and $\sigma_{\textmd{P},x\Delta}$ (and the corresponding vanilla option strikes) necessitates a continuous representation of volatilities across strikes, as the vanilla option strikes do not correspond to the market strangle strikes. Let $\sigma(K)$ be such a representation, we have then the following system to solve:

\begin{align}
	\sigma(K_{\textmd{ATM}}) &= \sigma_{\textmd{ATM}}\,,\label{eqn:atm-condition}\\
\Delta(1, K_{\textmd{C},x\Delta}, T, \sigma(K_{\textmd{C},x\Delta})) &= +x\Delta \%\,,\label{eqn:vanilla-delta-condition1}\\
\Delta(-1, K_{\textmd{P},x\Delta}, T, \sigma(K_{\textmd{P},x\Delta})) &= -x\Delta \%\,,\label{eqn:vanilla-delta-condition2}\\
V(-1,K_{\textmd{MS},\textmd{P},x\Delta}, T, \sigma(K_{\textmd{MS},\textmd{P},x\Delta}))
+ V(1,K_{\textmd{MS},\textmd{C},x\Delta}, T, \sigma(K_{\textmd{MS},\textmd{C},x\Delta})) &= V_{\textmd{MS},x\Delta} \,,\label{eqn:ms-condition}\\
\sigma(K_{\textmd{C},x\Delta}) -  \sigma(K_{\textmd{P},x\Delta}) &= \sigma_{\textmd{RR},x\Delta}	 \,. \label{eqn:risk-reversal-condition}
\end{align}
Due to the difference between $K_{\textmd{MS},\textmd{P},x\Delta}$ and $K_{\textmd{P},x\Delta}$, there are nine different strikes involved, not just five. The strikes $K_{\textmd{C},x\Delta}, K_{\textmd{P},x\Delta}$ may be solved exactly using Equations \ref{eqn:vanilla-delta-condition1} and  \ref{eqn:vanilla-delta-condition2} for given volatilities. We then have 5 remaining equations to fit the volatility representation $\sigma(K)$. There are several ways to perform the fit.

\citet{clark2011foreign} describes how to solve the system assuming a single unknown, the smile butterfly volatility $\sigma_{\textmd{BF},25\Delta}$, and does not try to fit to $10\Delta$ quotes. We adapt the algorithm to handle $10\Delta$ and $25\Delta$ quotes together:
\begin{enumerate}
	\item Choose an initial guess for the smile butterfly volatilities $\{\sigma_{\textmd{BF},25\Delta},\sigma_{\textmd{BF},10\Delta}\}$, typically the market butterfly quotes $\{\bar\sigma_{\textmd{BF},25\Delta},\bar\sigma_{\textmd{BF},10\Delta}\}$ (which corresponds to slightly different strikes).
	\item Use Equation \ref{eqn:vanilla-from-bf} to obtain $\sigma_{\textmd{C},25\Delta},\sigma_{\textmd{P},25\Delta},\sigma_{\textmd{C},10\Delta},\sigma_{\textmd{P},10\Delta}$ from the risk reversal quotes. Equation \ref{eqn:vanilla-from-bf} is exact when the vanilla volatilities correspond to the smile implied $\Delta$s, which is the case here.
	\item Calculate the vanilla option strikes by solving Equations \ref{eqn:vanilla-delta-condition1} and  \ref{eqn:vanilla-delta-condition2}.
	\item Find parameters of $\sigma(K)$ using a least-squares-optimizer on the volatilities $\sigma_{\textmd{ATM}}, \sigma_{\textmd{C},25\Delta},\sigma_{\textmd{P},25\Delta},\sigma_{\textmd{C},10\Delta},\sigma_{\textmd{P},10\Delta}$.
	\item Price up the market strangle using $\sigma(K)$. The least-squares error in model price against the market price (Equation \ref{eqn:ms-condition}) defines the objective.
\end{enumerate}
A Gauss-Newton optimizer \cite{klare2013gn} may be used to find the least-squares solution in option prices. The unknown $\sigma_{\textmd{BF},25\Delta}$ may be negative, it is perfectly acceptable \citep{wystup2018butterfly}. The bound constraint is for the vanilla volatilities given by Equation \ref{eqn:vanilla-from-bf} to be positive.

This presupposes that the representation $\sigma(K)$ fits well to the ATM, 10$\Delta$ and 25$\Delta$ strikes: the smile fitting procedure to vanillas is somewhat decoupled from the fitting to market strangles. For non-exact interpolations, it may be particularly relevant to add the error in ATM and RR volatilities, corresponding to Equations \ref{eqn:atm-condition} and \ref{eqn:risk-reversal-condition}, to the objective in step (5). The output has then a dimension of five. The intent is to recouple the vanilla fit with the market strangle fit. There is however a caveat: the market strangle error is minimized in terms of price, while the ATM and RR errors are minimized in terms of volatility. In order to obtain a similar error scale, an inverse vega weight $w_{\textmd{MS},x\Delta}$ must be used in the market strangle model vs. market price error: 
\begin{equation}
	w_{\textmd{MS},x\Delta} = \frac{1}{\left[ \phi\left(d_1(K_{\textmd{MS},\textmd{P}},\sigma_{\textmd{ATM}}+\bar\sigma_{\textmd{BF},x\Delta})\right) + \phi\left(d_1(K_{\textmd{MS},\textmd{C}},\sigma_{\textmd{ATM}}+\bar\sigma_{\textmd{BF},x\Delta})\right)\right] \sqrt{T} F(0,T)}\,.
\end{equation}

An additional subtlety is how and when to solve for the strikes of the $\Delta$ based quotes. Indeed, the two risk-reversals define vanilla $\Delta$-based quotes according to the currency pair and maturity convention. In Step 3, \citet{clark2011foreign} suggests that the smile parameterization should be used, and thus Step 4 should actually be merged with Step 3, such that we solve for the strike, using the smile. This makes a difference if the objective of dimension 2, and we will adopt this approach in our numerical examples, but it should not matter much if the objective is of dimension 5 with an error in RR volatilities computed according to the 10$\Delta$ and 25$\Delta$ vanilla options, as implied by the calibrated smile in Step 4.  

Finally, it is also possible to perform a single minimization, where the parameters of the representation $\sigma(K)$ are the unknown: the initial guess would be computed based on the smile convention, the objective would be the extended objective of dimension 5 based on the volatilities given by the parameters. It is in general less practical, since this direct algorithm does not allow for an easy reuse of an existing vanilla option smile calibration, while the above algorithm is identical regardless of the smile interpolation. The outcome of the calibration should be the same as the nested algorithm when the extended objective of dimension 5 is also used there.

In this article, we examine a few counterexamples where the fitting procedure does not necessary lead to an adequate interpolation of vanilla option prices.

\section{At-the-money error}
We consider options of maturity 147 days on EUR/HKD as of January 25, 2024 (Table \ref{tbl:eurhkd_147d}).
With exact interpolations such as a spline in log-moneyness, a cubic spline in forward $\Delta$ or a polynomial in $\Delta$, the vanilla implied volatilities are almost exactly the same. The SABR approximation\footnote{In the case of FX options it is usual to set the SABR parameter $\beta=1$ to avoid any probability mass at zero.} of \citet{hagan2002managing} does not fit exactly, which leads to small differences in the vanilla implied volatilies.
\begin{table}[H]
	\caption{Options on EUR/HKD expiring in 147 days as of January 25, 2024. $F(T)=8.500504$, $B_{\textmd{EUR}}(T)=0.9848102$, spot = 8.510111.\label{tbl:eurhkd_147d}}
	\centering{\begin{tabular}{ccccc}\toprule
	ATM &25$\Delta$-RR & 25$\Delta$-BF & 10$\Delta$-RR & 10$\Delta$-BF \\
	6.575 & -0.647 & 0.202 & -1.200 & 0.570  \\\bottomrule
	\end{tabular}}
\end{table}

\begin{figure}[H]
	\centering{
		\includegraphics[width=\textwidth]{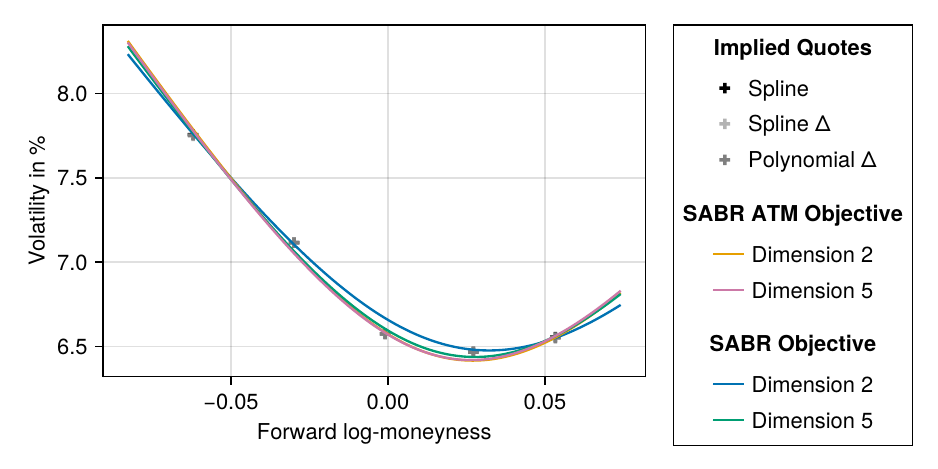}}
	\caption{Vanilla implied volatilities for EUR/HKD 147 days options with the SABR parameterization.\label{fig:eurhkd_sabr}}
\end{figure}

There are two potential issues with the final objective of dimension 2,  which consists only in the error in strangle prices:
\begin{itemize}
\item The ATM error may be larger than we wish. Figure\ref{fig:eurhkd_sabr} shows an absolute ATM error in volatility of 0.09\%, which dominates the other errors.
 A remedy is to match exactly the ATM volatility, thus reducing the number of parameters to fit, which is a common SABR calibration practice \citep{west2005calibration}.
\item The error in risk-reversals is not controlled. 
\end{itemize}



\begin{figure}[H]
	\centering{
		\includegraphics[width=\textwidth]{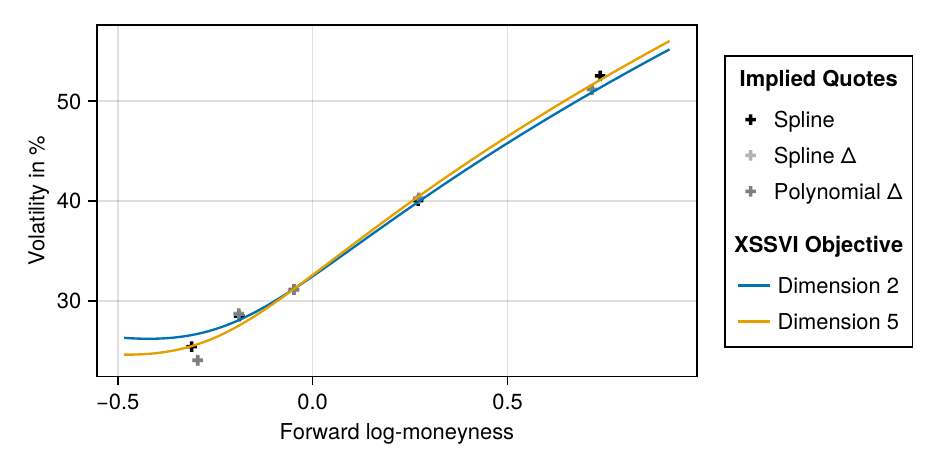}}
	\caption{Vanilla implied volatilities for EUR/TRY 1y options with the XSSVI parameterization.\label{fig:eurtry_xssvi}}
\end{figure}

\begin{table}[H]
 	\caption{Absolute error in weighted MS prices and RR, ATM vols \%. Violet indicates errors not part of the final objective.\label{tbl:eurtry_error}}
 	\centering{
 	\begin{tabular}{lllccccc}\toprule
	Maturity &	Model & Obj. dim. & 25$\Delta$ MS & 10$\Delta$ MS & 25$\Delta$ RR  & $10\Delta$ RR & ATM \\\midrule
	1y & XSSVI & 2&	0.0864 & -0.0367 &  \textcolor{violet}{0.1629} & \textcolor{violet}{-2.8797} & \textcolor{violet}{0.0552}\\
	& & 5&0.0869 & -0.3024 & 1.2725 & -0.6790 & 0.0471\\	
	2y& & 2 & 0.0000 & 0.0000 & 2.7042 & 1.7177& 0.0422\\
	&&5 & -0.2498 & -0.2881 & 1.3137 & -0.6915& 0.1043\\\bottomrule	

	\end{tabular}}
 \end{table}
Figure \ref{fig:eurhkd_sabr} displays the implied volatility smiles corresponding to the various calibration strategies. 
The RR calibration issue is clearer when calibrating the XSSVI parameterization of \citet{corbetta2019robust} to EUR/TRY options of maturity 1 year as of 2022/11/29 (Figure \ref{fig:eurtry_xssvi}). 
Using the extended objective of dimension 5 proves to be key here.

Looking at those makes us think we may fit SABR or XSSVI directly to the vanilla option volatilities implied by an exact interpolation such as a cubic spline in log-moneyness from the ATM, RR and BF quotes. This is a third approach to the FX smile calibration.  Its main advantages would be simplicity, robustness (because the interpolations are exact, and thus the conversions with market convention deltas are more direct), and easy reuse of existing model calibration algorithms to vanillas.
In the next section we will discover that the choice of exact interpolation may matter more than we would like.

\section{When exact interpolations lead to different volatilities}
We consider now options on EUR/TRY of maturity 2 year as of 2022/11/29 (Table \ref{tbl:eurtry_1y}). The Turkish Lira is particularly volatile, due to the high inflation in Turkey in 2022 and the resulting implied volatility smile has a somewhat unusual shape. 
\begin{table}[H]
	\caption{Quotes for options on EUR/TRY as of 2022/11/29 in forward $\Delta$ with premium.
	 For the maturity of 1 year, we have $r_{\textmd{TRY}} = 37.73\%$, $r_{\textmd{EUR}}=1.784\%$, spot=19.3483.
	 For the maturity of 2 years, we have $r_{\textmd{TRY}} = 37.65\%$, $r_{\textmd{EUR}}=1.950\%$. The $\star$ indicates manufactured quotes.
	 \label{tbl:eurtry_1y}}
	\centering{\begin{tabular}{cccccc}\toprule
	Maturity & ATM &25$\Delta$-RR & 25$\Delta$-BF & 10$\Delta$-RR & 10$\Delta$-BF \\
	6m & 22.12 & 9.385 & 2.187 & 21.148 & 7.633\\
1y	& 31.13 & 11.568 & 2.931 & 27.120 & 9.307 \\\cmidrule(lr){2-6}
	&	10$\Delta$-Put& 25$\Delta$-Put & ATM & 25$\Delta$-Call & 10$\Delta$-Call \\
1y	& 24.08 & 28.64 & 31.13 & 40.21 & 51.20 \\\bottomrule
	\end{tabular}}
\end{table}

We calibrate the cubic spline in log-moneyness, cubic spline in forward $\Delta$ and polynomial in simple $\Delta$. They interpolate exactly the vanilla option volatilities and thus, the objective dimension does not matter. It turns out that the resulting vanilla implied volatilities can be quite different, especially the 10$\Delta$ Put and Call, even though the error in MS, RR and ATM prices is zero. This is also an example where the simple smile convention approximation does not lead to correct vanilla implied volatilities. 
The exponential polynomial in simple\footnote{$\Delta_{\textmd{simple}}=\Phi(\log(F/K)/\sigma\sqrt{T})$. Contrary to Clark, we use the actual volatility and fully solve the strike for a given $\Delta$.} $\Delta$ of \cite{clark2011foreign} leads to nearly the same implied vanilla quotes as the cubic spline on forward $\Delta$.
\begin{figure}[H]
	\centering{
		\includegraphics[width=\textwidth]{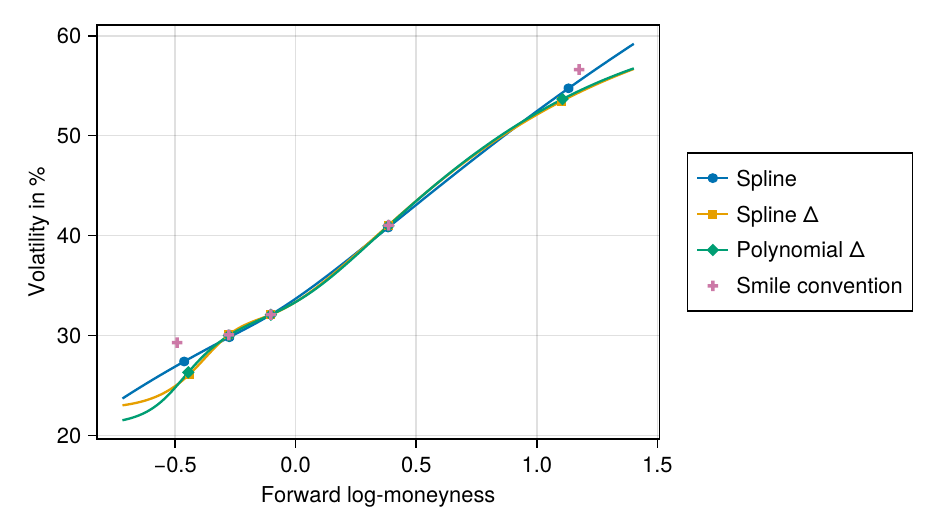}}
	\caption{Exact interpolations calibrated to options on EUR/TRY expiring in 2 year as of 2022/11/29.\label{fig:eurtry_2y_exact}}
\end{figure} 

\begin{figure}[H]
	\subfloat[SABR]{
		\includegraphics[width=0.5\textwidth]{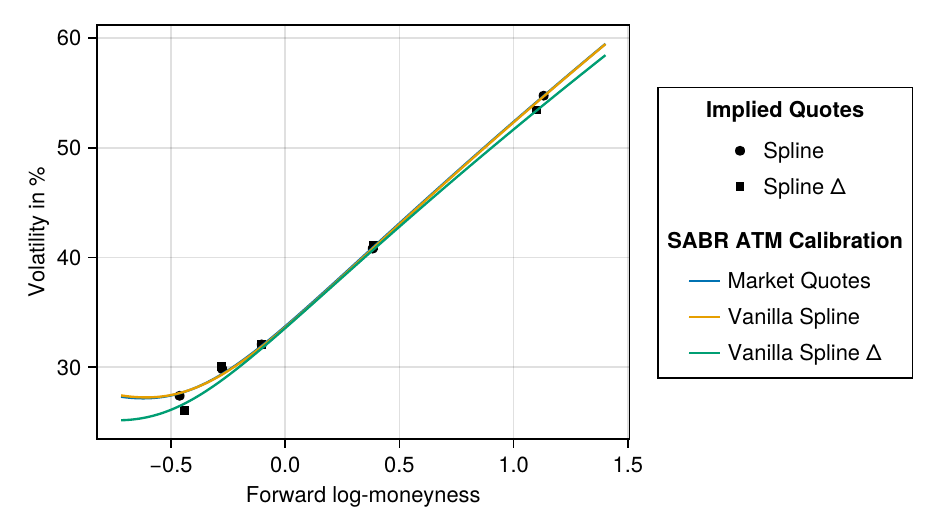}}
	\subfloat[XSSVI]{
			\includegraphics[width=0.5\textwidth]{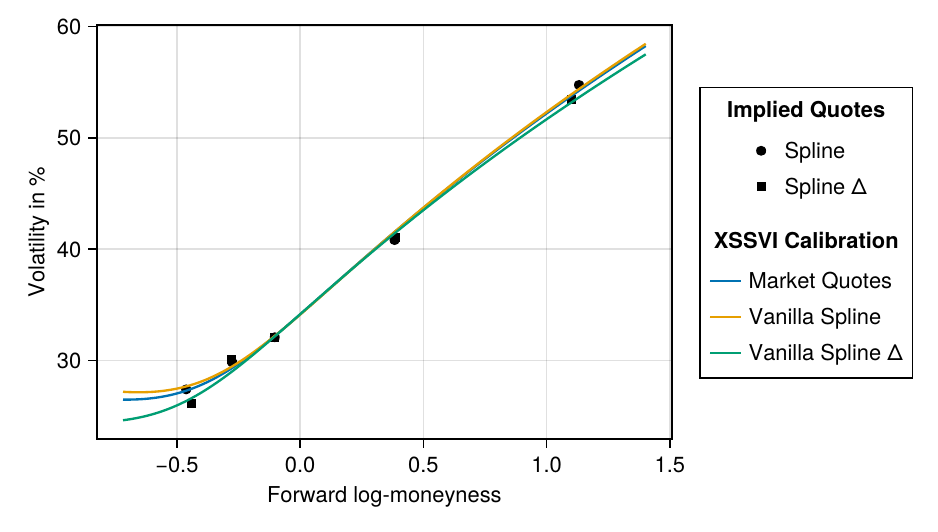}}	
	\caption{SABR and XSSVI calibrated to vanilla implied quotes for EUR/TRY expiring in 2 year as of 2022/11/29. \label{fig:eurtry_2y_sabr}}
\end{figure}

Which vanillas are better to calibrate to? In Figure \ref{fig:eurtry_2y_sabr}, we calibrate SABR and XSSVI to the vanilla option volatilities implied by the cubic spline in $\Delta$ and in log-moneyness. On this somewhat extreme data, the full model calibration is very close to the calibration of vanillas implied by the log-moneyness based spline.

\begin{figure}[H]
	\centering{
		\includegraphics[width=0.7\textwidth]{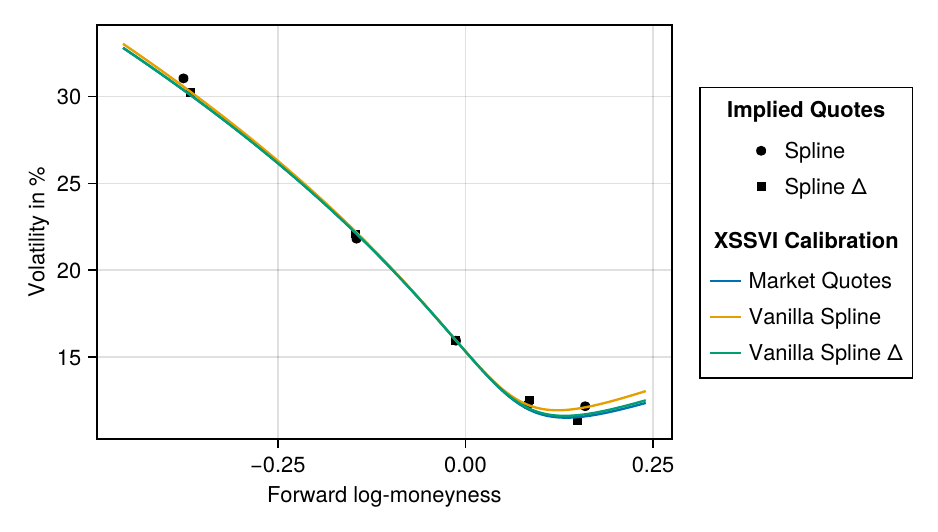}}
	\caption{Exact interpolations calibrated to options on USD/JPY expiring in 1 year.\label{fig:usdjpy_1y_xssvi}}
\end{figure}

This is not always the case, in Figure \ref{fig:usdjpy_1y_xssvi}, we calibrate XSSVI to options on USD/JPY of maturity 1 year with quotes given in  \citep[Table 3.4 and Section 3.5.5]{clark2011foreign}. The Delta based spline calibration is then the closest to the full model calibration. Here, the SABR calibration (not displayed) is nearly identical regardless of the choice of approach. 


\section{Manufactured example}
\begin{figure}[H]
	\centering{
		\includegraphics[width=0.7\textwidth]{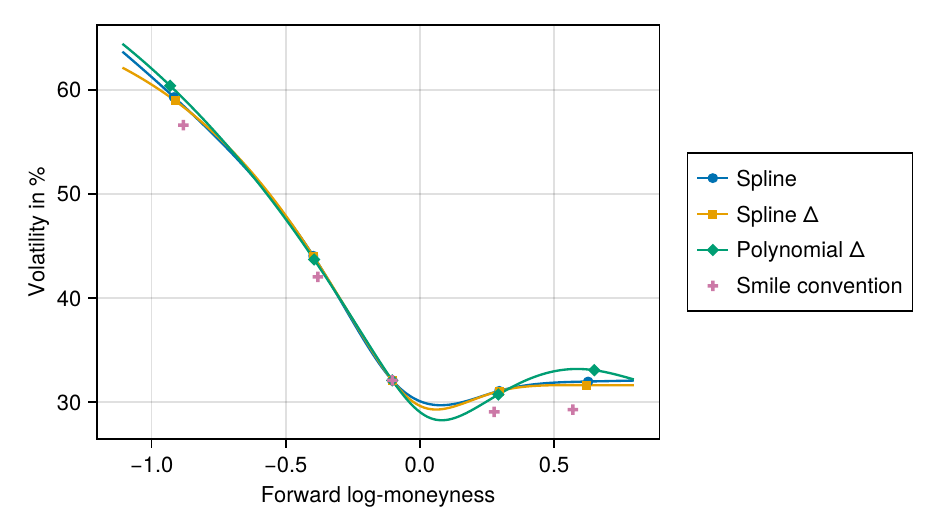}}
	\caption{Exact interpolations calibrated to manufactured quotes for options on EUR/TRY options expiring in 2 year as of 2022/11/29. The polynomial in $\Delta$ oscillates.\label{fig:eurtry_2yman_exact}}
\end{figure}

If we bump slightly the $25\Delta$-RR quote and change the sign of the RR quotes in Table \ref{tbl:eurtry_1y}, we still have arbitrage-free quotes.
The calibration against those manufactured market quotes  using a polynomial in $\Delta$ leads to an oscillating smile.
The resulting vanilla implied volatilities at $10\Delta$ are also clearly distinct from the ones of the spline based calibrations (Figure \ref{fig:eurtry_2yman_exact}).

\section{Numerical issues}
\subsection{Call $\Delta$ with premium}
Delta lookups can be tricky:
the Call $\Delta_{F,\textmd{pct}}$ is a non monotonic function of strike, or equivalently, of log-moneyness. 
As a consequence, \citet{reiswich2012fx} advise to first search for the maximum and then solve for the strike on the right side of the maximum, corresponding to OTM strikes. \citet{jackel2020strike} proposes a fast, accurate and robust numerical method tailored to this problem.

\begin{figure}[H]
	\centering{
		\includegraphics[width=0.7\textwidth]{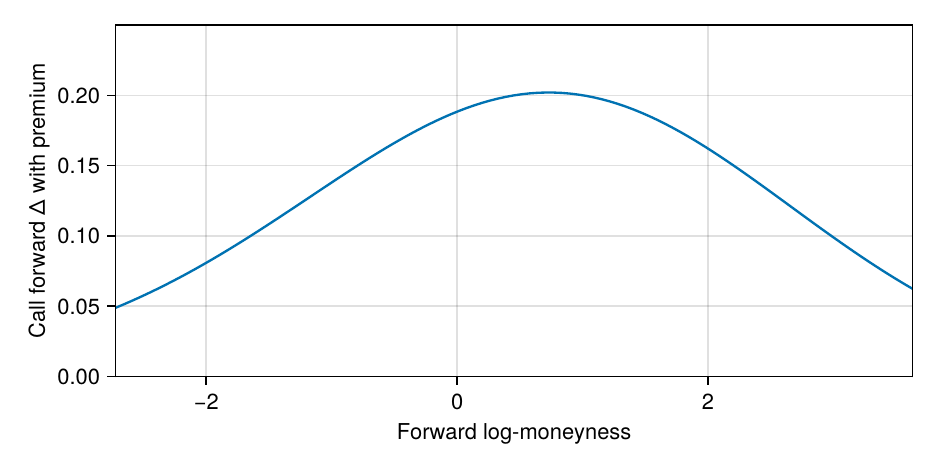}}
	\caption{Call Forward $\Delta$ with premium as a function of log-moneyness for a large volatility $\sigma=125\%$ and $T=2$ years.
	Notice the non-monotonicity and the peak not reaching $\Delta=25\%$.\label{fig:calldeltawithpremium}}
\end{figure}

It becomes even more challenging if we calibrate the smile function based on the market Delta quotes directly, which is essentially  what is described in \citep{clark2011foreign}. A non-linear solver needs to be used to lookup the model volatility at given $\Delta$s, for each choice of model parameters done by the smile calibration minimizer.
Some $\Delta$s may not be reachable by any strike as, for high volatilities, the maximum $\Delta_{F,\textmd{pct}}$ can be below 25\% (e.g. Figure \ref{fig:calldeltawithpremium}). 
This means that the smile function may not be able to produce a quote for the 25$\Delta$ call option.
There are other edge cases, such as very low quoted vols, or exploding model vols, which may happen in the wings of SABR and lead to non-sensical Deltas, during the minimization.

\subsection{Delta for a given strike}
Some smile representations we considered involve a direct $\Delta$ based interpolation.
For example, the polynomial in $\Delta$ function with coefficients $(a_i)_{i=0,...,5}$ reads
\begin{equation}
	\sigma(\Delta_{\textmd{simple}}) = \exp\left(\sum_{i=0}^4 a_i \Delta_{\textmd{simple}}(K,\sigma)^i\right)\label{eqn:poly_delta}
\end{equation}
where $\Delta_{\textmd{simple}}(K,\sigma) = \Phi\left(\frac{\ln F(0,T)/K}{\sigma \sqrt{T}}\right)$.

\begin{figure}[H]
	\subfloat[Smile in $\Delta_{\textmd{simple}}$]{
		\includegraphics[width=0.49\textwidth]{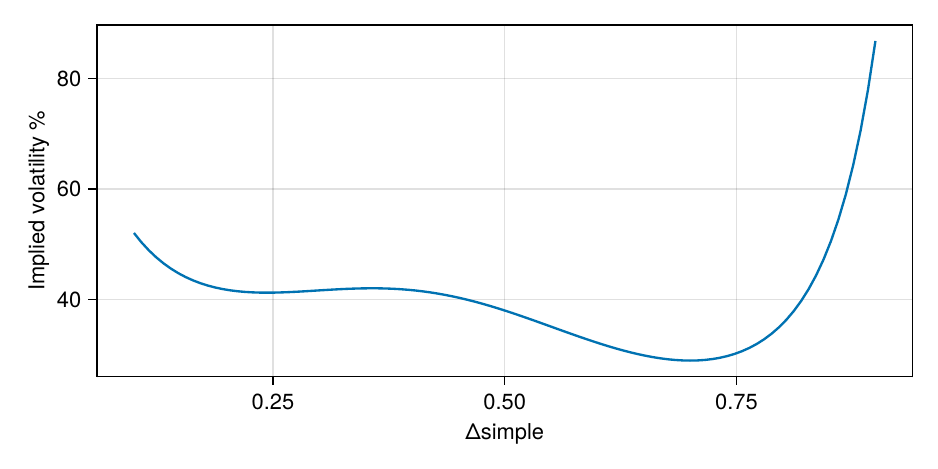}}
	\subfloat[$\Delta_{\textmd{simple}}(K,\sigma)$ is monotonic]{
			\includegraphics[width=0.49\textwidth]{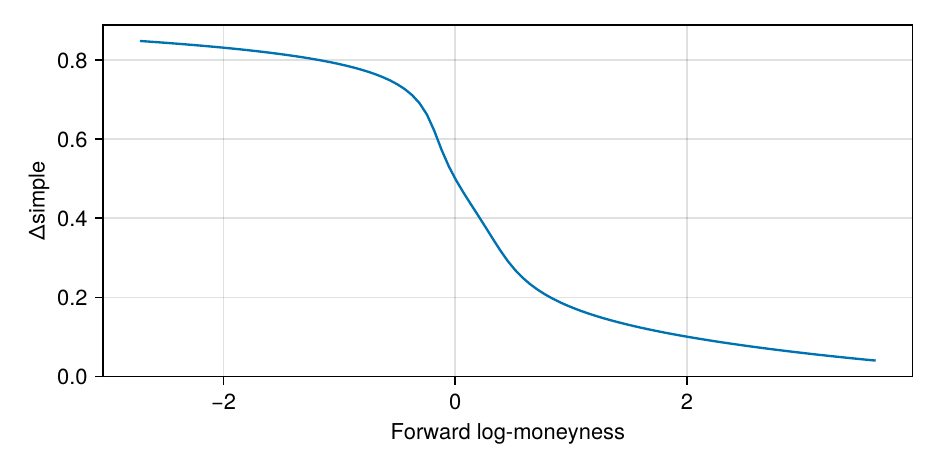}}	
	\caption{Example of problematic  polynomial in $\Delta_{\textmd{simple}}$. \label{fig:polypb}}
\end{figure}

In order to price vanilla options, we need to find the volatility for a given fixed strike, with the above implicit definition of 
the smile function in terms of $\Delta_{\textmd{simple}}$.
\citet[Section 1.4.9]{wystup2017fx} describes the simple fixed point iteration:
\begin{enumerate}
	\item Choose $\sigma_0$ = at-the-money vol.
	\item Calculate $\Delta_{n+1} = \Delta_{\textmd{simple}}(K,\sigma_n)$.
	\item Take $\sigma_{n+1} = \sigma(\Delta_{n+1})$, using Equation \ref{eqn:poly_delta}.
	\item If $|\sigma_{n+1}-\sigma_n| < \epsilon$, then quit, otherwise continue with Step 2.
\end{enumerate}
This leads to an infinite loop for $\bm{a}=(0.114, -11.8, 49.2, -84.1, 48.5)$, $T=2$, $F(0,T)=39.51$, $\sigma_0=32\%$, $K=10$.
The smile corresponding to this polynomial is presented in Figure \ref{fig:polypb}.
A better idea would be to use the Newton's method:
\begin{enumerate}
	\item Choose $\sigma_0$ = at-the-money vol.
	\item Define the objective function $f(v)= \sigma(\Delta_{\textmd{simple}}(K,v)) - v$, using Equation \ref{eqn:poly_delta}.
	\item Use Newton's method on $f$, starting with $v=\sigma_0$.
\end{enumerate}
This also fails, this time for a lower strike, for example $K=5$.
A more robust way is to define the objective as a function of $\Delta$ instead of volatility:
\begin{enumerate}
	\item Define the objective function $g(d)= \Delta_{\textmd{simple}}(K,\sigma(d)) -d$, using Equation \ref{eqn:poly_delta}.
	\item Use bracketing solver such the TOMS748 solver of \citet{alefeld1995algorithm} to solve $g(d)=0$ on [0,1]. This leads to a solution $d_{\textmd{final}}$.
	\item Then $\Delta_{\textmd{simple}}(K,\sigma)=d_{\textmd{final}}$.
\end{enumerate}
If, instead of a bracketing solver, we were to use Newton's method, $d$ may pushed beyond the range [0,1].
It is possible to rely on Newton's method, using a transformation $h$ on $d$ from $\mathbb{R} \to (0,1)$, such as the sigmoid function.

\subsection{Non monotonic Deltas}
\begin{figure}[H]
	\subfloat[Smile]{
		\includegraphics[width=0.5\textwidth]{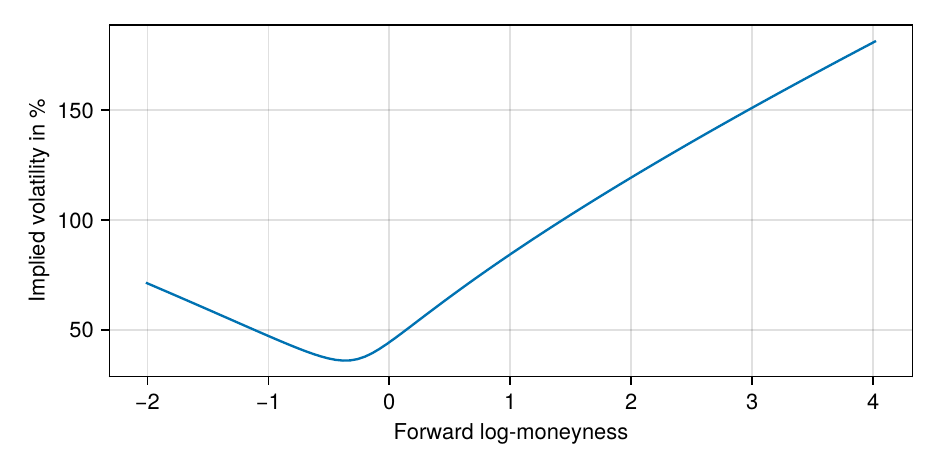}}
	\subfloat[Call $\Delta_{F,\textmd{pips}}$]{
			\includegraphics[width=0.5\textwidth]{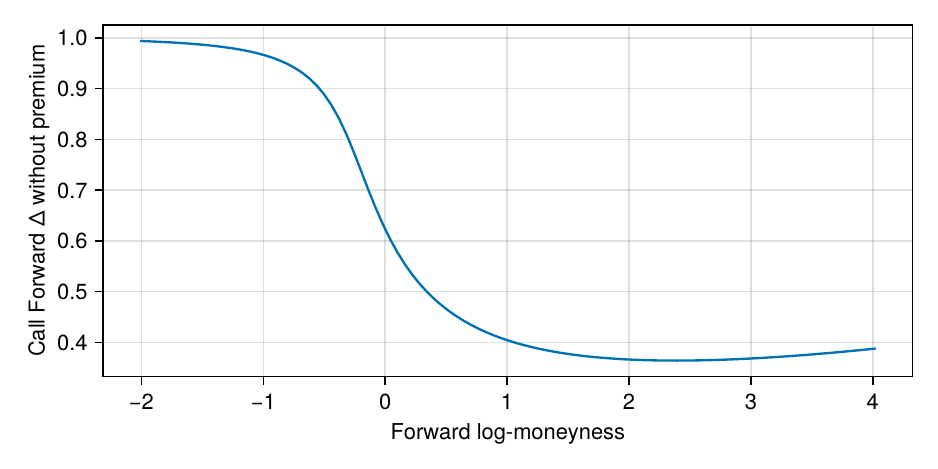}}	
	\caption{A problematic SABR smile. The  $\Delta_{F,\textmd{pips}}$ does not reach 25\% and is not monotonic.\label{fig:sabr_bad}}
\end{figure}

During the calibration of SABR, some parameters tried by the minimizer may lead to arbitrages, because we use only an approximation for the SABR model (the one from \citet{obloj2007fine}), a common practice, but this expansion is really only valid for a small vol-of-vol. A consequence is a non monotonic forward $\Delta$, even for Put-$\Delta$ or $\Delta_{F,\textmd{pips}}$ (without premium). Two examples of SABR parameters are provided in Table \ref{tbl:sabr_bad}. Figure \ref{fig:sabr_bad} shows the non-monotonic Call $\Delta_{F,\textmd{pips}}$, which does not reach 30\%, meaning the SABR wing is bad, and has arbitrages.

\begin{table}[H]
	\caption{Examples of SABR parameters which lead to non monotonic $\Delta_{F,\textmd{pips}}$ with $F(0,T)=39.512$, $T=2$, $B(0,T)=0.96175$.\label{tbl:sabr_bad}}
	\centering{\begin{tabular}{lccccc}\toprule
		Set & $\alpha$ & $\beta$ & $\rho$ & $\nu$\\\midrule
		I & 34.9\% & 1 &54.4\% & 127\%\\
		II & 37.5 \% & 1 & 66.0\% & 100\%\\\bottomrule
	\end{tabular}}
\end{table}
Those parameters do not allow to obtain a price for a $10\Delta$ or a $25\Delta$ Call. This will need to be dealt with in not-so-obvious manners in the minimization. A remedy may be to use an arbitrage-free implementation of SABR \citep{lefloch2014finite}.

\section{Conclusion}
We reviewed different ways to calibrate a model to the market quotes, and how this choice may impact the outcome of the calibration.

The most precise is to fully calibrate directly the model to the market MS, RR and ATM quotes, but care needs to be taken such that the final objective is sensitive to all quotes. There is then a negligible difference between a calibration of the smile parameterization directly to volatilities in the market $\Delta$ convention or to volatilities in fixed strikes. While calibrating directly to quotes in $\Delta$s is attractive, is suffers from the eventuality of many more numerical issues, and requires more computational resources.

The simplest, and likely more robust way is to use an exact interpolation to imply some vanilla implied volatilities, which replicate exactly the market quotes. And then to calibrate the chosen smooth parameterization on top of those. The exact interpolation  bypasses many caveats with $\Delta$ inversions.
We have seen that there is no unique solution to this problem, the use of a cubic spline on log-moneyness seems the simplest. The difference in the error against market quotes is usually negligible (Table \ref{tbl:calibration_error}).

\acknowledgments{The author would like to thank Dr. Gary Kennedy and Dr. Fabien Le Floc'h for fruitful conversations and feedback on early drafts of this paper.}
\externalbibliography{yes}
\bibliography{arbfree_interpolation.bib}

@article{alefeld1995algorithm,
  title={Algorithm 748: Enclosing zeros of continuous functions},
  author={Alefeld, GE and Potra, Florian A and Shi, Yixun},
  journal={ACM Transactions on Mathematical Software (TOMS)},
  volume={21},
  number={3},
  pages={327--344},
  year={1995},
  publisher={ACM New York, NY, USA}
}

@article{lefloch2014finite,
  title={Finite difference techniques for arbitrage free SABR},
  author={{Le Floc'h} and Kennedy, Gary J},
  journal={Journal of Computational Finance},
  volume={20},
  number={3},
  pages={51--79},
  year={2017}
}

@article{obloj2007fine,
  title={Fine-tune your smile: Correction to Hagan et al},
  author={Obl{\'o}j, Jan},
  journal={arXiv preprint arXiv:0708.0998},
  year={2007}
}

@article{jackel2020strike,
  title={Strike from volatility and delta-with-premium},
  author={J{\"a}ckel, Peter},
  journal={Quantitative Finance},
  volume={20},
  number={8},
  pages={1227--1235},
  year={2020},
  publisher={Taylor \& Francis}
}

@article{klare2013gn,
	title={GN--a Simple and Effective Nonlinear Least-Squares Algorithm for the Open Source Literature.},
	author={Klare, Kenneth and Miller, Guthrie},
	year={2013},
	url={http://www.netlib.org/misc/gn/}
}

@article{wystup2018butterfly,
	title={Negative Butterflies and Why We Check Butterfly Arbitrage by a Non-Negative Probability Density},
	author={Wystup, Uwe},
	journal={Wilmott},
	volume={2022},
	number={121},
	pages={48--49},
	year={2022},
	publisher={John Wiley \& Sons, Ltd Chichester, UK}
}

@article{hagan2002managing,
	title={Managing smile risk},
	author={Hagan, Patrick S and Kumar, Deep and Lesniewski, Andrew S and Woodward, Diana E},
	journal={Wilmott magazine},
	year={2002}
}

@book{clark2011foreign,
  title={Foreign exchange option pricing: A practitioner's guide},
  author={Clark, Iain J},
  year={2011},
  publisher={John Wiley \& Sons}
}

@article{reiswich2012fx,
  title={FX volatility smile construction},
  author={Reiswich, Dimitri and Wystup, Uwe},
  journal={Wilmott},
  volume={2012},
  number={60},
  pages={58--69},
  year={2012},
  publisher={Wiley Online Library}
}

@article{corbetta2019robust,
  title={Robust calibration and arbitrage-free interpolation of {SSVI} slices},
  author={Corbetta, Jacopo and Cohort, Pierre and Laachir, Ismail and Martini, Claude},
  journal={Decisions in Economics and Finance},
  volume={42},
  number={2},
  pages={665--677},
  year={2019},
  publisher={Springer}
}

@book{wystup2017fx,
  title={FX options and structured products},
  author={Wystup, Uwe},
  year={2017},
  publisher={John Wiley \& Sons}
}

@article{west2005calibration,
  title={Calibration of the SABR model in illiquid markets},
  author={West, Graeme},
  journal={Applied Mathematical Finance},
  volume={12},
  number={4},
  pages={371--385},
  year={2005},
  publisher={Taylor \& Francis}
}
\appendixtitles{no}

\appendix
\section{Calibration errors}
\begin{table}
	\caption{Calibration error corresponding to the objective of dimension 5, for different option sets, using a calibration, (a) on top of vanilla quotes implied by a spline in log-moneyness, (b) on top of vanilla quotes implied by a spline in $\Delta$, (c1) with the objective of dimension 2 and converting strikes up-front before the smile calibration step 4, (c2) with the objective of dimension 2 and steps 3 and 4 merged such that the parameterization is directly calibrated to $\Delta$ quotes, (d1) with the objective of dimension 5 and converting strikes up-front before the smile calibration step 4, (d2) with the objective of dimension 5 and steps 3 and 4 merged.\label{tbl:calibration_error}}
	\centering{
\begin{tabular}{lllrr}\toprule
	Option set & Model & Calibration & \multicolumn{2}{c}{$\ell_2$ Error}\\\cmidrule(lr){4-5}
	& & & Fixed Strikes & Fixed $\Delta$s\\
	USD/AED 7d & ATM SABR & Spline & 0.00005&\\
	           &          & Spline $\Delta$ & 0.00005&\\
	           &          & Dimension 2 &  0.00258 & 0.00004\\
			   &          & Dimension 5 & 0.00004 & 0.00004\\
	           & SABR     & Dimension 5 & 0.00004 & 0.00004\\
			   & XSSVI    & Spline & 0.00259&\\
			   &          & Spline $\Delta$ & 0.00261&\\
			   &          & Dimension 2 & 0.00491 &0.00491 \\
			   &          & Dimension 5 & 0.00149 &0.00149 \\\cmidrule(lr){1-5}
    EUR/TRY 2y & ATM SABR & Spline & 0.00679&\\
	&          & Spline $\Delta$ &0.01140&\\
	&          & Dimension 2 &  0.00821 & 0.00831\\
	&          & Dimension 5 & 0.00665 & 0.00663\\
	& SABR     & Dimension 5 & 0.00665 & 0.00663\\
	& XSSVI    & Spline & 0.01557&\\
	&          & Spline $\Delta$ & 0.01657&\\
	&          & Dimension 2 & 0.01593 & 0.03203 \\
	&          & Dimension 5 & 0.01536 &0.01536 \\\cmidrule(lr){1-5}
	EUR/HKD 147d   & ATM SABR & Spline &0.00069&\\
	&          & Spline $\Delta$ &0.00069&\\
	&          & Dimension 2 & 0.00722 & 0.00073\\
	&          & Dimension 5 & 0.00068 & 0.00067\\
	& SABR     & Dimension 5 & 0.00059 & 0.00059\\
	& XSSVI    & Spline & 0.00057&\\
	&          & Spline $\Delta$ & 0.00057&\\
	&          & Dimension 2 &  0.01870 & 0.00083 \\
	&          & Dimension 5 & 0.00056 &0.00057 \\\cmidrule(lr){1-5}
AUD/NZD 1w & ATM SABR & Spline &0.00219&\\
&          & Spline $\Delta$ &0.00219&\\
&          & Dimension 2 & 0.00238 & 0.00215\\
&          & Dimension 5 &0.00215 & 0.00215\\
& SABR     & Dimension 5 &0.00218 & 0.00214\\
& XSSVI    & Spline & 0.00212&\\
&          & Spline $\Delta$ & 0.00212&\\
&          & Dimension 2 &  0.00213 & 0.00211 \\
&          & Dimension 5 & 0.00211 &0.00211  \\\cmidrule(lr){1-5}
USD/JPY 1y & ATM SABR & Spline &0.00325&\\
&          & Spline $\Delta$ &0.00314&\\
&          & Dimension 2 & 0.00280 & 0.00266\\
&          & Dimension 5 &0.00260 & 0.00257\\
& SABR     & Dimension 5 &0.00260 & 0.00257\\
& XSSVI    & Spline &0.00856&\\
&          & Spline $\Delta$ & 0.00771&\\
&          & Dimension 2 &  0.00876 & 0.00749 \\
&          & Dimension 5 & 0.00737&0.00743  \\\bottomrule
\end{tabular}}
\end{table}
	
\end{document}